\documentclass[1L2pt,preprint]{aastex}

\begin{document}

\title{On the Star Formation Law for Spiral and Irregular Galaxies}

\author{Bruce G. Elmegreen}
\affil{IBM Research Division, T.J. Watson Research Center, 1101 Kitchawan Road,
Yorktown Heights, NY 10598, bge@us.ibm.com}

\begin{abstract}
A dynamical model for star formation on a galactic scale is proposed in which the
interstellar medium is constantly condensing to star-forming clouds on the dynamical time
of the average midplane density, and the clouds are constantly being disrupted on the
dynamical time scale appropriate for their higher density. In this model, the areal star
formation rate scales with the 1.5 power of the total gas column density throughout the
main regions of spiral galaxies, and with a steeper power, 2, in the far outer regions and
in dwarf irregular galaxies because of the flaring disks.  At the same time, there is a
molecular star formation law that is linear in the main and outer parts of disks and in
dIrrs because the duration of individual structures in the molecular phase is also the
dynamical time scale, canceling the additional 0.5 power of surface density. The total gas
consumption time scales directly with the midplane dynamical time, quenching star formation
in the inner regions if there is no accretion, and sustaining star formation for $\sim100$
Gyr or more in the outer regions with no qualitative change in gas stability or molecular
cloud properties. The ULIRG track follows from high densities in galaxy collisions.
\end{abstract}


\section{Introduction}
\label{intro}

The increase of the star formation rate (SFR) per unit area, $\Sigma_{\rm SFR}$, with
increasing molecular gas surface density, $\Sigma_{\rm mol}$, is close to linear in the
main parts of galaxy disks \citep{bigiel08,leroy08,bigiel11,schruba11}. The usual
interpretation is that star formation occurs in molecular clouds at a fixed rate per
molecule, independent of environment \citep[e.g.,][]{krum09}. However, molecular clouds
should be denser in the inner parts of galaxies because of larger pressures, as shown by
\cite{heyer15} for the Milky Way, and denser clouds should form stars faster.  A more
physical interpretation is that stars always form at the local ISM collapse rate regardless
of the molecular fraction. This would give a slope of 1.5 in a plot of $\Sigma_{\rm SFR}$
versus total gas column density, $\Sigma_{\rm gas}$, in the inner regions where the disk
scale height is about constant, and a slope of 2 for this relation in the outer regions
where there is a disk flare (see below). The observed linear relation for molecules would
then be the result of an additional correlation between the local dynamical time of
molecular gas and its duration in that phase. This is the interpretation we investigate
here.

Numerical simulations are of limited value in determining the underlying physics of the
areal star formation law because they need to make assumptions about what happens inside
their equivalent of molecular clouds. \cite{gnedin14} showed that the areal law reflects
the local (3D) law. That is, a local dynamical law like $\rho_{\rm SFR}\propto\rho_{\rm
gas}^{3/2}$ for 3D star formation and gas densities $\rho_{\rm SFR}$ and $\rho_{\rm gas}$
translates into a large-scale law like $<\rho_{\rm SFR}>\propto<\rho_{\rm gas}>^{3/2}$ for
average densities, and into an areal law $\Sigma_{\rm SFR}\propto\Sigma_{\rm gas}^{3/2}$.
The implication is that the linear star formation law for molecules, which is often
described as a constant consumption time for molecules on a kpc scale, corresponds to a
constant time per molecule inside each molecular cloud.  The threshold model by
\cite{lada10} and \cite{evans14}, as observed in the solar neighborhood, is an example of
such a law because there is a constant threshold column density, and presumably also
density, for star formation.

A problem with this model is that at the high pressures of inner galaxy disks, clouds of a
given column density are less strongly self-gravitating than they are in the solar
neighborhood. Recall that the influence of self-gravity on a cloud, as opposed to external
pressure, $P$, is directly proportional to $G\Sigma_{\rm cloud}^2/P$. For a given cloud
surface density $\Sigma_{\rm cloud}$, self-gravity becomes exponentially less important at
higher $P$ \citep{padoan12}. Thus there can be neither a constant threshold column density
for star formation nor a linear proportion between the SFR and the molecular cloud mass
over galactic scales where the equilibrium pressure from the weight of the gas layer varies
by a large amount. In galaxy disks, the pressure varies approximately as the square of the
surface density, which corresponds to a factor of $\exp(8)=3000$ over 4 scale lengths,
i.e., over most of the main disk.

\cite{bonnell13} simulated cloud formation, cloud collisions, and collapse in the denser
parts of clouds throughout a spiral arm. They found the observed steep relation
$\Sigma_{\rm SFR}\propto\Sigma_{\rm gas}^{1.4}$ \citep{kennicutt98} for the total gas
$\Sigma_{\rm gas}$ at the same time as there was a linear relation with the dense gas,
$\Sigma_{\rm SFR}\propto\Sigma_{\rm gas,dense}$, as observed for CO-emitting gas
\citep{bigiel08}. They attributed the difference to cloud formation in the low density
phase, with an increase in the dense gas fraction from cloud collisions as the total gas
density increased.  Here we take a similar point of view, but at a more fundamental level.

We propose that most of the ISM is in a constant state of collapse to star formation
followed by dispersal from young stellar pressures
\citep[e.g.,][]{li05,bournaud10,henne14}. The duration of each state and its probability
are proportional to the associated dynamical time.  Star formation is part of this cycle
and has a rate given by the same principle, i.e., a fixed fraction of ambient gas turns
into stars during its dynamical time. This model is in agreement with observations of cloud
phase lifetimes and turbulent crossing times \citep[e.g.,][]{elmegreen00}. We show here
that the model also gives the observed molecular and total-gas SFR laws with their slopes
of 1 and 1.5, respectively.

There are many models for the SFR-surface density relation. \cite{krum05} proposed a
dynamical model with star formation at locally unstable densities in a log-normal density
pdf; in \cite{renaud12}, this density threshold is where the gas is cool enough to be
supersonically turbulent. \cite{krum09} extended the model to spiral galaxies assuming a
constant surface density for individual molecular clouds in main disks because of pressure
decoupling from the ambient medium, and an increasing surface density for individual clouds
in the inner parts where the ambient pressure is high. \cite{krumholz12} made the same
assumptions, extending the high pressure regime to high redshift and interacting galaxies.
HI-dominated galaxies like dwarf irregulars were considered in \cite{krumholz13}.
\cite{ostriker10} considered feedback-controlled pressure for two stable HI phases with the
excess gas in self-gravitating CO clouds that have a constant SFR/mass. \cite{kim13} and
\cite{kim15} ran simulations with star formation feedback that explain the scale height and
properties of the ISM with a dynamically-based SFR at a threshold density and a constant
efficiency per unit free fall time. \cite{hu15} use the dynamical model for SFR in dwarf
galaxies with a constant threshold density and efficiency for star formation. In
\cite{krum09}, \cite{dib11}, \cite{murante15} and elsewhere, the SFR depends explicitly on
the $H_2$ fraction that comes from molecule formation theory. \cite{hopkins14} simulate
galaxies with star formation at the local collapse rate in the molecular parts of
self-gravitating clouds that exceed a certain fixed density, noting that the rate does not
depend sensitively on the detailed star formation law or prescription for molecule
formation.

Most of these and other models share attributes of the proposal here. Key differences are
that we do not need an equilibrium basic state, although there is a mass balance between
collapse and break-up of star-forming clouds, we do not identify molecular clouds with star
forming clouds, or vice versa, and we do not assume standardized molecular clouds with a
constant star formation rate per molecule (that constant results from the model
automatically). The most similar previous theory was that expressed by \cite{madore10}, who
considered star formation on a timescale equal to the sum of an interstellar collapse time
and a cloud break up time. We agree that timing is important, and take the next logical
step in using this timing argument to convert the physically motivated SFR relation for
total gas into a shallower apparent SFR relation for the molecular part of the gas.

\section{The Dynamical Model for Star Formation}
\label{dynmodel}

The dynamical model for star formation \citep{larson69,madore77,silk87,katz92} suggests
that
\begin{equation}
\Sigma_{\rm SFR}=\epsilon_{\rm ff}\Sigma_{\rm gas}/t_{\rm ff}
\label{eq1}
\end{equation}
where
\begin{equation}
1/t_{\rm ff}=(32G\rho/[3\pi])^{1/2}=(16G\Sigma_{\rm gas}/[3\pi  H])^{1/2}.
\label{freefall}
\end{equation}
The second step uses $\rho=\Sigma_{\rm gas}/2H$. Combined, they give
\begin{equation}
\Sigma_{\rm SFR}=\epsilon_{\rm ff}(16G/[3\pi H])^{1/2}\Sigma_{\rm gas}^{3/2}.
\label{combined}
\end{equation}
In the inner parts of galaxies, $H$ may be about constant \citep{heyer15}. Setting $H=100$ pc for the total gas
layer and $\epsilon=0.01$ \citep{krum07}, we get
\begin{equation}
{{\Sigma_{\rm SFR}}\over{M_\odot\;{\rm pc}^{-2}\;{\rm Myr}^{-1}}}=
8.8\times10^{-5}\left({{\Sigma_{\rm gas}}\over{M_\odot\;{\rm pc}^{-2}}}\right)^{1.5}.
\label{eq:inner}
\end{equation}
This is the same as the relation observed for many types of galaxies by \cite{kennicutt12}, as shown in Figure
1a.

The efficiency per unit free fall time, $\epsilon_{\rm ff}$, should depend on cloud
geometric structure, i.e., the probability distribution function of dense gas
\citep{elmegreen02}, and feedback, i.e., the prevention of local collapse by the stars that
have already formed \citep[see review in][]{padoan14}. First-principle derivations of this
efficiency by several groups are compared in \cite{fed12}, who find they give about the
same value in various situations. We assume simply that $\epsilon_{\rm ff}$ is constant for
all galaxies \citep[see also][]{krumholz12}. We do not consider the average molecular
fraction to be a large part of the efficiency: diffuse (non-self-gravitating) gas can be
molecular and molecular gas, even at high excitation density, can be diffuse.  Similarly,
atomic clouds can be diffuse, but also self-gravitating if they are massive enough, which
gives them a low average density at a given pressure.

In the outer parts of spiral galaxies and in most of dwarf irregular galaxies, the gas dominates the disk mass
surface density and
\begin{equation}
H=\sigma^2/(\pi G\Sigma_{\rm gas}).
\label{eq:flare}
\end{equation}
Then
\begin{equation}
1/t_{\rm ff}=4G\Sigma_{\rm gas}/[3^{1/2}\sigma]
\label{eq:tff}
\end{equation}
and
\begin{equation}
\Sigma_{\rm SFR}=\epsilon_{\rm ff}(4/\sqrt{3})G\Sigma_{\rm gas}^2/\sigma .
\end{equation}
For a typical outer-disk and dIrr gas velocity dispersion of $\sigma=6$ km s$^{-1}$, and for $\epsilon_{\rm
ff}=0.01$ again, we obtain
\begin{equation}
{{\Sigma_{\rm SFR}}\over{M_\odot\;{\rm pc}^{-2}\;{\rm Myr}^{-1}}}
=1.7\times10^{-5}
\left({{\Sigma_{\rm gas}}\over{M_\odot\;{\rm
pc}^{-2}}}\right)^2 .
\label{eq:outer}
\end{equation}
This is approximately the same as the relation observed for 20 dIrr's in \cite{eh15} and
for 5 dIrr's and the outer parts of 17 spirals in \cite{bigiel10} as shown in Figure 1b.
The similarity between dIrr and outer spirals was also noted by \cite{roy15} who point out
further that this implies the SFR relation is independent of metallicity. Other regions
that are gas dominated, as may be the case for parts of mergers or interacting galaxies,
could have this $\Sigma_{\rm gas}^2$ relation locally as well.

Equations (\ref{eq:inner}) and (\ref{eq:outer}) suggest that the dynamical model with a
constant $\epsilon_{\rm ff}$ reproduces the superlinear SFR relation in the inner disks of
starburst galaxies where the slope is $\sim1.5$, as identified by \cite{kennicutt98}, and
it reproduces the relation in dIrrs and the outer parts of spiral galaxies where the slope
is $\sim2$ \citep{bigiel10, eh15}. The steeper slope is because of the disk flare
\citep[see also][]{barnes12}. Both of these regions have their total gas dominated by the
ISM phase that is actually observed, so the dynamical model is revealed directly. When
using the molecular surface density on the abscissa, however, we need to consider something
else. In the starburst inner region, the gas is mostly molecular, and in the far outer
regions of spirals and in dIrrs, the gas is mostly atomic. In the region between, the
molecular fraction varies with radius and this is where the linear molecular KS relation
appears. Thus the linear relation is most likely a by-product of a changing molecular
fraction. We discuss this in the next section.

\section{The Surface Density of CO}
\label{surfdensiity}

The dynamical model cannot have a linear relation between $\Sigma_{\rm SFR}$ and total gas
surface density because the space density inside clouds varies with ambient pressure in a
galaxy, and therefore with radius \citep[e.g.,][]{kalberla09} and spiral arm phase. The
linear relation between $\Sigma_{\rm SFR}$ and molecular surface density as measured by CO
\citep{bigiel08,leroy08} or HCN \citep{gao04,wu05} should therefore arise from an
additional effect. We suggest that all ISM structures evolve on their internal dynamical
timescales, which means the turbulent crossing time for both gravitating and diffuse clouds
and the gravitational time for self-gravitating clouds \citep{elmegreen00}. As a result,
molecular gas should evolve on a timescale approximately equal to the dynamical time at its
threshold density for excitation. The assumption here is that the average density of a
molecular cloud or subregion in that cloud is proportional to the excitation density of the
molecule observed \citep{krumholzthompson07,nara08}. This is reasonable because for
centrally condensed clouds, a high fraction of the mass has a density close to the value at
the observed edge. In addition, for power-law density structure inside clouds
\citep{lombardi15}, the ratio of gas masses for two fixed densities, such as a collapse
threshold and an excitation density, is constant.

In the dynamical model, gas is never in equilibrium but is either collapsing toward star
formation at the dynamical rate of the ambient medium, or getting disrupted after star
formation, which takes place on some multiple of the dynamical time scale at the density of
the star-forming cloud. For the life cycle of CO clouds, the first rate is $\sim1/t_{\rm
ff}$ evaluated at the average local midplane density, as above, and the second is
$\sim1/t_{\rm ff,CO}$ evaluated at the characteristic density of CO emission. Then the
fraction of the gas mass in a CO-emitting state is proportional to the mean time spent
there,
\begin{equation}
f_{\rm CO}\sim {{t_{\rm ff,CO}}\over{t_{\rm ff}+t_{\rm ff,CO}}}.
\end{equation}
Consequently, the average column density of CO is
\begin{equation}
\Sigma_{\rm CO}\sim f_{\rm CO}\Sigma_{\rm gas}.
\label{sigmaco1}
\end{equation}
These timescales are proportional to the square roots of the densities for our assumption
of gravitationally-driven processes, so
\begin{equation}
f_{\rm CO}=(1+[\rho_{\rm CO}/\rho_{\rm gas}]^{1/2})^{-1}.
\label{fgrav}
\end{equation}
\cite{madore10} also considered gas phase timing as part of an explanation for the
SFR-density relation; what we call $t_{\rm ff,CO}$ for the star-forming gas, he calls a
stagnation time, which is the same basic concept.  What is new here is the identification
of this time with the dynamical time at the excitation threshold density for tracer
molecules like CO and the role that this time plays in converting an intrinsic 1.5 power of
total density or column density for the SFR to an observed 1.0 power for the molecules.

For most of a normal galaxy disk, $\rho_{\rm CO}>>\rho_{\rm gas}$ and $f_{\rm CO}\sim
(\rho_{\rm gas}/\rho_{\rm CO})^{1/2}$.  If CO emission occurs at a characteristic density
for CO excitation, then $\rho_{\rm CO}$ is about constant and $f_{\rm CO}\propto \rho_{\rm
gas}^{1/2}\propto 1/t_{\rm ff}$.  In this limit,
\begin{equation}
\Sigma_{\rm CO}=\left({{\rho_{\rm gas}}\over{\rho_{\rm CO}}}\right)^{1/2}\Sigma_{\rm gas} =
{{\Sigma_{\rm SFR}}\over{\epsilon_{\rm ff}}}\left({{3\pi}\over{32G\rho_{\rm CO}}}\right)^{1/2}
\label{eq:sigmaco}
\end{equation}
which may be inverted to give
\begin{equation}
\Sigma_{\rm SFR}=\epsilon_{\rm ff}\Sigma_{\rm CO}/t_{\rm ff,CO}.
\end{equation}
Here $t_{\rm ff,CO}$ uses $\rho_{\rm CO}$ instead of $\rho$ in equation (\ref{eq:tff}) and
is constant because $\rho_{\rm CO}$ is approximately the constant excitation density.

We have reproduced the Bigiel-Leroy linear molecular law by stretching the physical $\Sigma_{\rm gas}^{3/2}$ law
toward lower $\Sigma_{\rm gas}$ as the SFR drops when plotting versus $\Sigma_{\rm CO}$ because the molecular
fraction is dropping at the same time.

Equations (\ref{eq1}), (\ref{freefall}), (\ref{sigmaco1}), and (\ref{fgrav}) may be combined to give the SFR for
arbitrary molecular transitions,
\begin{equation}
\Sigma_{\rm SFR}=\epsilon_{\rm ff}\Sigma_{\rm mol}\left(1+[\rho_{\rm gas}/ \rho_{\rm
mol}]^{1/2}\right)(32G\rho_{\rm mol}/[3\pi])^{1/2}.
\label{combined}
\end{equation}
where $\rho_{\rm mol}$ is some factor of order unity times the excitation density. The
ambient density enters equation(\ref{combined}) because the SFR in the KS relation is
usually taken to be the average over some large region of a galaxy, where the average
density has this value. This relation makes a smooth transition from a $3/2$ power law
dependence on molecular gas surface density in the far-inner regions where $\rho_{\rm
mol}\sim\rho_{\rm gas}$ and $\rho_{\rm gas}\sim\Sigma_{\rm gas}/2H$ for constant $H$ (cf.
Eq. \ref{eq:inner}), to a linear dependence on molecular gas at intermediate and large
radii (independent of $H$) where $\rho_{\rm mol}>>\rho_{\rm gas}$.  Because molecules are
not usually observed at large radius, the equation reverts to the quadratic dependence on
the total $\Sigma_{\rm gas}$ in the outer flaring region (cf. Eq. \ref{eq:outer}). Note
that for either CO or HCN in the main disk, $\rho_{\rm mol}>>\rho_{\rm gas}$ and the SFR is
$\epsilon_{\rm ff}\Sigma_{\rm mol}/t_{\rm ff,mol}$ for constant $t_{\rm ff,mol}$ at some
factor times the appropriate excitation density. For all of these relations, $\epsilon_{\rm
ff}\sim$ constant in principle, although at higher densities, such as for HCN,
$\epsilon_{\rm ff}$ should be larger than it is for CO because there is less gas structure
closer to the thermal scale, where gas motions approach the thermal speed
\citep{elmegreen08}.

\section{$\Sigma_{\rm SFR}$ for the Kennicutt and Bigiel-Leroy correlations}
\label{together}

Figure 2 shows the superlinear and linear laws in the normal way they are plotted, one with $\Sigma_{\rm SFR}$
versus total gas surface density and the other with $\Sigma_{\rm SFR}$ versus CO surface density. We assume a
model galaxy with an exponential radial profile of gas surface density \citep[e.g.,][]{kalberla09,heyer15}
\begin{equation}
\Sigma_{\rm gas}=\Sigma_{\rm gas,0}\exp{\left(-r/r_{\rm D}\right)}.
\end{equation}
The disk scale height is assumed to be constant, $H_0$ for $\Sigma_{\rm gas}>10\;M_\odot$
pc$^{-2}$ where stars tend to dominate gas in spiral galaxies, and flaring as
$H=H_0(10\;M_\odot\;{\rm pc}^{-2}/\Sigma_{\rm gas})$ outside this radius (cf. eq.
\ref{eq:flare}), where gas tends to dominate stars. These assumptions are based on
observations, as noted above, and should be quantified further as part of a more complete
theory.

The left-hand plot is the radial profile of the SFR. We assume four cases for illustration,
$\Sigma_{\rm gas,0}=10^2$, $10^3$ and $10^4$ in $M_\odot$ pc$^{-2}$ for $H_0=100$ pc and
$\Sigma_{\rm gas,0}=10^3\;M_\odot$ pc$^{-2}$ for $H_0=10$ pc as in a thin nuclear disk. The
radial profiles are double exponential because the flare in the outer part decreases the
dynamical rate with radius more rapidly than in the inner part. The middle panel shows the
Kennicutt relation with total gas surface density on the abscissa. All cases with the same
$H_0$ converge to the same relation so the curves lie on top of each other. The slope is
1.5 in the inner region and 2 in the outer region. The right-hand panel shows the molecular
law. In the inner regions (the ``starburst region'') the slope is 1.5 because most of the
ISM is molecular. At intermediate and large radii the slope is unity because the CO gas has
a spatially varying mass fraction that is directly proportional to the ambient dynamical
rate. Also in the right-hand plot are dashed lines that show the total gas $\Sigma_{\rm
gas}$ repeated from the middle panel.  The horizontal distance between the solid and dashed
lines is the molecular fraction. These theoretical relationships reproduce the basic
observations without requiring that star formation follows molecule formation or that
star-forming clouds are standardized.

\section{Implications and Conclusions}
\label{conclusions}

We have shown that a model in which all ISM structures continuously evolve on a dynamical
time with star formation following dynamical contraction and cloud dispersal following star
formation reproduces the superlinear relation between $\Sigma_{\rm SFR}$ and total gas
surface density, $\Sigma_{\rm gas}$, with a slope of $\sim1.5$. It also reproduces the
steeper relation in the far outer parts of spiral galaxies and in dwarf irregular galaxies
where the disks flare. At the same time, it reproduces the linear relation between the
areal SFR and CO surface density in the main disks of spirals because of a varying
molecular mass fraction. These observed relations have been obtained without assuming that
molecular clouds are standardized or have constant SFRs per unit mass, or that molecular
clouds have a universal threshold column density for star formation. Such standardization
may apply locally, but it seems unlikely over large regions of a galaxy disk given the
enormous pressure range.

Because of this intrinsic dependence of the SFR per unit gas mass on the gas density, the
dynamical model predicts that the consumption time of total gas varies inversely with the
square root of density, i.e., with $\Sigma_{\rm gas}^{-1/2}$ and $\Sigma_{\rm SFR}^{-1/3}$
over regions of galaxies where the scale height is about constant. Similarly, the local
consumption time varies as $\Sigma_{\rm gas}^{-1}$ and $\Sigma_{\rm SFR}^{-1/2}$ in
gas-dominated regions. As a result of these inverse variations, the inner parts of galaxies
should be exhausted of gas first, quenching their star formation if cosmic accretion of
fresh gas cannot readily get to the inner regions in present-day galaxies. For the same
reasons involving density, star formation in starbursts, mergers, and ULIRGs should end
much more rapidly than in the main disks of normal galaxies. We also obtain a consumption
time of $\sim100$ Gyr seen by \cite{bigiel10} and \cite{eh15} in far-outer disks and dIrrs
without requiring any special assumptions about molecule formation or stabilization.

The similarity in the KS relation between local galaxies and the star forming population at
intermediate redshifts, $z\sim1-3$, is the result of a similar average gas density in this
interpretation. While the SFRs and column densities of high-$z$ galaxies can be $\sim10$
times higher than in local galaxies, the thicknesses of high-$z$ disks are $\sim10\times$
higher too, as they are forming the thick-disk components of modern galaxies
\citep{elmegreen06,bournaud09,comeron14}. This constancy of $\rho_{\rm gas}$ is a sensible
result if the Toomre parameter $Q\sim\kappa \sigma/[\pi G \Sigma_{\rm gas}]$ is regulated
by disk instabilities because then for a given galaxy type, i.e., a given mass, size, and
therefore average epicyclic frequency $\kappa$, we have $\sigma\propto\Sigma_{\rm gas}$,
which gives a disk thickness $H= \sigma^2/[\pi G\Sigma_{\rm gas}]\propto \sigma
\propto\Sigma_{\rm gas}$, and therefore a density $\rho_{\rm gas}=\Sigma_{\rm
gas}/[2H]\sim$ constant. The ULIRG track in the KS relation is also the same at high and
low redshift \citep{tacconi13,sargent14,silverman15,bethermin15} because in both cases it
follows from the elevated average density that results from disk-disk collisions and
tidally-driven gas inflows \citep{teyssier10}.

The presence of a change in the slope of the KS relation at $\Sigma_{\rm
gas}\sim10\;M_\odot$ pc$^{-2}$ \citep{bigiel08,leroy08} is usually attributed to a sudden
change in the molecular fraction because individual molecular clouds lose their
self-shielding at about this column density \citep{krumholz09b}. While this molecular
transition is approximately true, there is another interpretation in the dynamical model
where the slope change results from the disk flare. In this interpretation, the flare
begins when the stellar surface density drops to less than the gas surface density and the
gas velocity dispersion stops decreasing significantly with radius. Using either equations
(\ref{eq:inner}) or (\ref{eq:outer}) for $\Sigma_{\rm gas}=10\;M_\odot$ pc$^{-2}$, we
derive $\Sigma_{\rm SFR}\sim10^{-3}\;M_\odot$ Myr$^{-1}$ pc$^{-2}$ at this position. Thus,
in a Hubble time of $10^4$ Myr, the stellar surface density also reaches $\sim10\;M_\odot$
pc$^{-2}$. We note that for both outer spiral disks or dIrr galaxies, there is no sudden
drop off in the SFR below this $\Sigma_{\rm gas}$, as would be expected if molecular clouds
suddenly lost their shielding ability \citep{eh15}. Shielding is important from an
observational point of view, however, as the transition from molecular-dominant to
atomic-dominant gas should still occur at a value of $\Sigma_{\rm gas}$ that scales
inversely with the metallicity \citep{krumholz09b}. The only point here is that in the
dynamical model, this molecular transition does not corresponds to a qualitative change in
star formation.

The dynamical model has several predictions. First, the consumption time that is derived for each different
molecular transition should scale approximately with the dynamical time at the excitation density of the
molecule. The scaling should not be exactly linear because the efficiency factor, $\epsilon_{\rm ff}$, should
also change for different molecules if we include additional effects such as depletion of the molecule onto
grains or a change in the gas substructure as the thermal scale for turbulence is approached.  A second
prediction is that a column density threshold for star formation that may appear in certain molecular transitions
should increase approximately as the square root of the ambient pressure. This is because clouds with high ratios
of boundary pressure to squared-column density are more stable against gravitational collapse than clouds with
low ratios.

I am grateful to M. Rafelski for pointing out the similarity in the KS relations for dIrrs
and outer spiral disks and to D.A. Hunter and P. Padoan for discussions.

\begin{figure}\epsscale{1.0}\plottwo{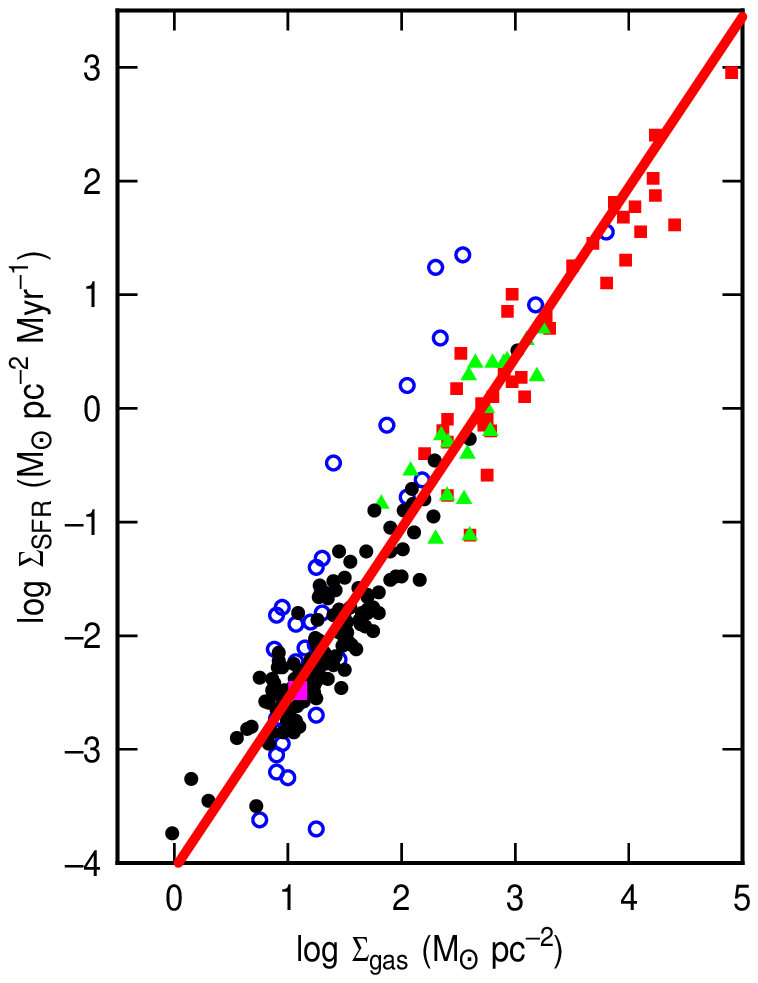}{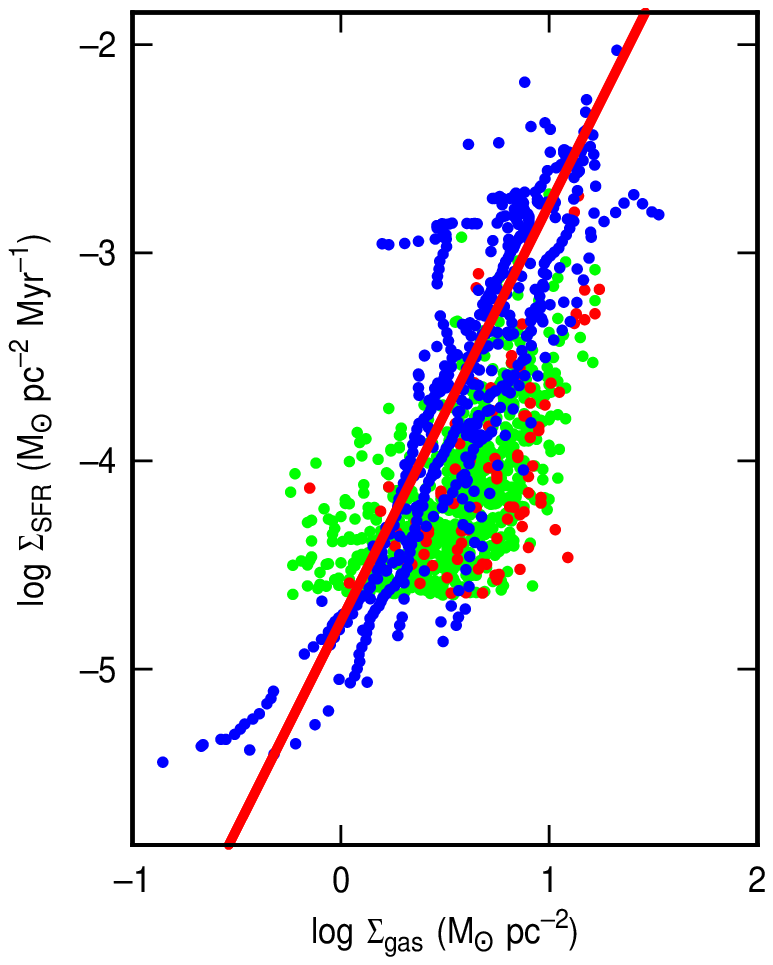}
\caption{(Left) The observed SFR -- gas surface density correlation integrated over galaxies
from \cite{kennicutt12} with equation (\ref{eq:inner}) overlayed as a red line (the plotted points were redrawn
from the original figure; the Milky Way is the magenta square, normal galaxies are black points,
blue circles are metal poor $[Z<0.3Z_\odot]$), red squares are IR selected, and
green triangles are circum-nuclear starbursts. (Right) The resolved
star formation relations in the far outer regions of spirals (red points) and dwarf Irregulars (green points)
from \cite{bigiel10} and in dIrrs (blue points) from \cite{eh15}. The theory from equation (\ref{eq:outer})
is overlayed as a red line. } \label{figure1}\end{figure}

\begin{figure}\epsscale{1.0}\plotone{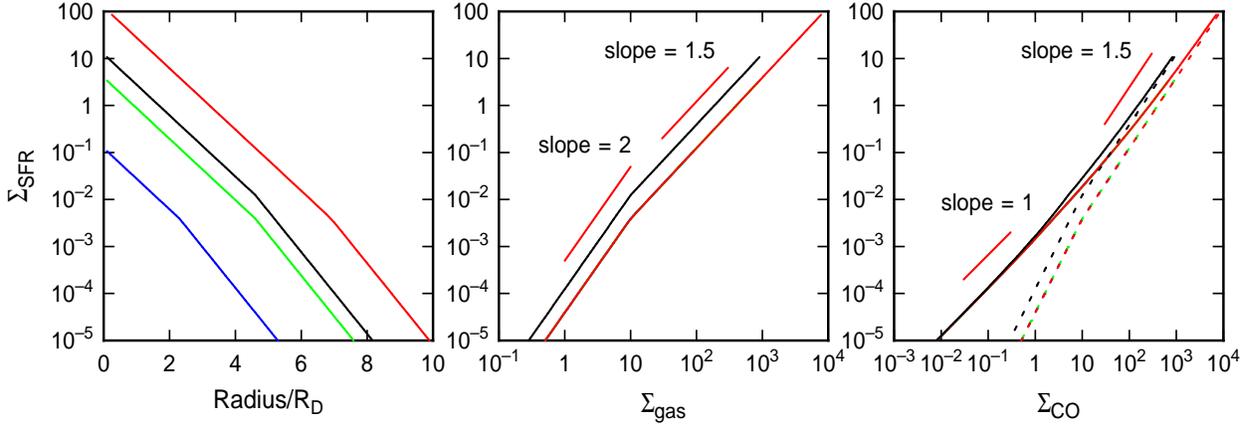}
\caption{Idealized examples of the dynamical model assuming an exponential radial profile of total
disk gas and a constant efficiency, $\epsilon_{\rm ff}=0.01$. Four cases are considered:
$\Sigma_{\rm gas,0}=10^2$ (blue), $10^3$ (green) and $10^4$ (red) in $M_\odot$ pc$^{-2}$ for scale height $H_0=100$ pc, and $\Sigma_{\rm
gas,0}=10^3\;M_\odot$ pc$^{-2}$ for $H_0=10$ pc (black). (Left) The radial profiles of star formation.
(Middle) Areal SFR versus total gas surface density, showing the 1.5 power law in the inner regions and the
2.0 power law in the outer regions. The blue, green and red curves lie on top of each other and are
not distinguishable. The black curve
corresponding to small thickness has a high SFR for its surface density, as in a ULIRG track, because of its high
midplane density. (Right) The molecular SFR showing the 1.5 power law
at very high column density, where most of the gas is molecular, and the 1.0 power law elsewhere where the molecular fraction
varies with the relative time spent in the dense cloud phase. The dashed curves repeat the total-gas curves in the middle panel. For
any $\Sigma_{\rm SFR}$, the horizontal distance between the solid and dashed lines is the molecular fraction.}
\label{figure2}\end{figure}

\begin{thebibliography}

\bibitem[Barnes et al.(2012)]{barnes12} Barnes, K.\ L., van Zee, L., C\^ot\'e, S., \&
    Schade, D. 2012, ApJ, 757, 64

\bibitem[B\'ethermin et al.(2015)]{bethermin15} B\'ethermin, M., Daddi, E., Magdis, G., et al.
    2015, A\&A, 573, A113

\bibitem[Bigiel et al.(2008)]{bigiel08} Bigiel, F., Leroy, A., Walter, F.,
    Brinks, E., de Blok, W. J. G., Madore, B., \& Thornley, M. D. 2008, AJ,
    136, 2846

\bibitem[Bigiel et al.(2010)]{bigiel10} Bigiel, F., Leroy, A., Walter, F., Blitz, L.,
    Brinks, E., de Blok, W.J.G., Madore, B. 2010, AJ, 140, 1194

\bibitem[Bigiel et al.(2011)]{bigiel11} Bigiel F. et al., 2011, ApJ, 730, L13

\bibitem[Bonnell et al.(2013)]{bonnell13} Bonnell, I.A., Dobbs, C.L., \& Smith, R.J. 2013, MNRAS, 430, 1790

\bibitem[Bournaud, Elmegreen \& Martig(2009)]{bournaud09} Bournaud, F., Elmegreen, B.G., \& Martig, M. 2009,
    ApJ, 707, L1

\bibitem[Bournaud et al.(2010)]{bournaud10} Bournaud, F., Elmegreen, B.G., Teyssier, R.,
    Block, D.L., \& Puerari, I. 2010, MNRAS, 409, 1088

\bibitem[Comer\'on et al.(2014)]{comeron14} Comer\'on, S., Elmegreen, B.G., Salo, H., Laurikainen, E.,
    Holwerda, B.W., Knapen, J.H. 2014, A\&A, 571, 58

\bibitem[Dib(2011)]{dib11} Dib, S. 2011, ApJ, 737, L20

\bibitem[Elmegreen(2000)]{elmegreen00} Elmegreen, B.G. 2000, ApJ, 530, 277

\bibitem[Elmegreen(2002)]{elmegreen02} Elmegreen, B.G. 2002, ApJ, 577, 206

\bibitem[Elmegreen(2008)]{elmegreen08} Elmegreen, B.G. 2008, ApJ, 672, 1006

\bibitem[Elmegreen \& Elmegreen(2006)]{elmegreen06} Elmegreen, B.G. \& Elmegreen, D.M. 2006, ApJ, 650, 644

\bibitem[Elmegreen \& Hunter(2015)]{eh15} Elmegreen, B.G., \& Hunter, D.A. 2015, ApJ, 805, 145

\bibitem[Evans et al.(2014)]{evans14} Evans, N.J., II., Heiderman, A., \&
    Vutisalchavakul, N. 2014, ApJ, 782, 114

\bibitem[Federrath \& Klessen(2012)]{fed12} Federrath, C., \& Klessen, R.S. 2012, ApJ, 761, 156

\bibitem[Gao \& Solomon(2004)]{gao04} Gao, Y., \& Solomon, P.M. 2004, ApJS, 152, 63

\bibitem[Genzel et al. (2014)]{genzel14} Genzel, R., F\"orster Schreiber, N. M., Lang, P., et al. 2014,
    ApJ, 785, 75

\bibitem[Gnedin et al.(2014)]{gnedin14} Gnedin, N.Y., Tasker, E.J., \& Fujimoto, Y. 2014,
    ApJL, 787, L7

\bibitem[Hennebelle \& Iffrig(2014)]{henne14} Hennebelle, P., \& Iffrig, O. 2014, A\&A,
    570, A81

\bibitem[Heyer \& Dame(2015)]{heyer15} Heyer, M., \& Dame, T.M. 2015, ARA\&A, 53, 583

\bibitem[Hopkins et al.(2014)]{hopkins14} Hopkins, P.F., Keres, D., O\~norbe, J.,
    Faucher-Gigu\`ere, C.-A., Quataert, E., Murray, N., \& Bullock, J.S. 2014, MNRAS, 445, 581

\bibitem[Hu et al.(2015)]{hu15} Hu, C.-Y., Naab, T., Walch, S., Glover, S.C.O., Clark, P.
    C. 2015, arXiv151005644

\bibitem[Kalberla \& Kerp (2009)]{kalberla09} Kalberla, P.M.W. \& Kerp, J. 2009, ARA\&A, 47, 27

\bibitem[Katz(1992)]{katz92} Katz, N. 1992 1992, ApJ, 391, 502


\bibitem[Kennicutt(1998)]{kennicutt98} Kennicutt, R.C., Jr. 1998, ApJ, 498, 541

\bibitem[Kennicutt \& Evans(2012)]{kennicutt12} Kennicutt R. C., \& Evans N. J., 2012, ARA\&A, 50, 531

\bibitem[Kim et al.(2013)]{kim13} Kim, C.-G., Ostriker, E.C., \& Kim, W.-T.
    2013, ApJ, 776, 1

\bibitem[Kim \& Ostriker(2015)]{kim15} Kim, C.-G., \& Ostriker, E.C. 2015, arXiv
    1511.00010

\bibitem[Krumholz \& McKee(2005)]{krum05} Krumholz, M. R., \& McKee, C. F. 2005, ApJ, 630,
    250

\bibitem[Krumholz \& Tan(2007)]{krum07}Krumholz, M. R., \& Tan, J. C. 2007, ApJ, 654, 304

\bibitem[Krumholz \& Thompson(2007)]{krumholzthompson07} Krumholz, M. R., \& Thompson, T.
    A. 2007, ApJ, 669, 289

\bibitem[Krumholz et al.(2009a)]{krum09} Krumholz, M. R., McKee, C. F., \& Tumlinson, J.
    2009, ApJ, 699, 850

\bibitem[Krumholz et al.(2009b)]{krumholz09b} Krumholz, M.R., McKee, C.F., \&
    Tumlinson, J. 2009, ApJ, 693, 216

\bibitem[Krumholz et al.(2012)]{krumholz12} Krumholz, M.R., Dekel, A., \& McKee,
    C.F. 2012, ApJ, 745, 69

\bibitem[Krumholz(2013)]{krumholz13} Krumholz, M.R. 2013, MNRAS, 436, 2747

\bibitem[Lada et al.(2010)]{lada10} Lada, C.J., Lombardi, M., \& Alves, J.F.
    2010, ApJ, 724, 687

\bibitem[Larson(1969)]{larson69} Larson, R.B. 1969, MNRAS, 145, 405

\bibitem[Leroy et al.(2008)]{leroy08} Leroy, A.K., Walter, F., Brinks, E., Bigiel, F.,
    de Blok, W. J. G., Madore, B., \& Thornley, M. D. 2008, AJ, 136, 2782

\bibitem[Li et al.(2005)]{li05} Li, Y., Mac Low, M.-M., \& Klessen, R.S.
    2005, ApJ, 620, L19

\bibitem[Lombardi et al.(2015)]{lombardi15} Lombardi, M., Alves, J., \& Lada, C.J. 2015, A\&A, 576, L1

\bibitem[Madore(1977)]{madore77} Madore, B. F. 1977, MNRAS, 178, 1

\bibitem[Madore(2010)]{madore10} Madore, B.F. 2010, ApJ, 716, L131

\bibitem[Murante et al.(2015)]{murante15} Murante, G., Monaco, P., Borgani,
    S., Tornatore, L., Dolag, K., Goz, D. 2015, MNRAS, 447, 178

\bibitem[Narayanan et al.(2008)]{nara08} Narayanan, D., Cox, T. J., Shirley, Y., et al.
    2008, ApJ, 684, 996

\bibitem[Ostriker et al.(2010)]{ostriker10} Ostriker, E.C., McKee, C.F., Leroy, A.K.  2010,
    ApJ, 721, 975

\bibitem[Padoan et al.(2012)]{padoan12} Padoan, P., Haugb\/olle, T., Nordlund, \AA., 2012, ApJ, 759, L27

\bibitem[Padoan et al.(2014)]{padoan14} Padoan, P., Federrath, C., Chabrier, G., Evans,
    N.J.,II, Johnstone, D., J\/orgensen, J. K., McKee, C. F., Nordlund, \AA. 2014, in
    Protostars and Planets VI, H. Beuther, R.S. Klessen, C.P. Dullemond,
    and T. Henning (eds.), Univ. Arizona Press, Tucson, p.77

\bibitem[Renaud et al.(2012)]{renaud12} Renaud, F., Kraljic, K., \& Bournaud, F. 2012, ApJ, 760, L16

\bibitem[Roychowdhury et al.(2015)]{roy15} Roychowdhury, S., Huang, M.-L., Kauffmann, G.,
    Wang, J., Chengalur, J.N. 2015, MNRAS, 449, 3700

\bibitem[Sargent et al.(2014)]{sargent14} Sargent, M. T., Daddi, E., Béthermin, M., et al. 2014, ApJ, 793, 19

\bibitem[Schruba et al.(2011)]{schruba11} Schruba, A., Leroy, A. K., Walter,
    F., et al. 2011, AJ, 142, 37
\bibitem[Silk(1987)]{silk87} Silk, J. 1987, IAUS, 115, 663

\bibitem[Silverman et al.(2015)]{silverman15} Silverman, J. D., Daddi, E., Rodighiero, G., et al.
    2015, ApJ, 812, L23

\bibitem[Tacconi et al.(2013)]{tacconi13} Tacconi, L. J., Neri, R., Genzel, R. et al. 2013, ApJ, 768, 74

\bibitem[Teyssier et al.(2010)]{teyssier10} Teyssier, R., Chapon, D., \& Bournaud, F. 2010, ApJ, 720, L149

\bibitem[Wu et al.(2005)]{wu05} Wu, J., Evans, N.J., II., Gao, Y., Solomon, P.M., Shirley, Y.L., \& Vanden Bout,
    P.A. 2005, ApJ, 635, L173


\end{thebibliography}
\end{document}